\documentclass[rnote,traditabstract]{aa} 
\usepackage{graphicx}
\usepackage{natbib}
\usepackage{txfonts}
\usepackage{color}
\newcommand{\pks}{PKS~0447-439}

\begin{document}
   \title{On the redshift of the blazar \pks\thanks{This paper includes data gathered with 
       the 6.5 meter Magellan Telescopes located at Las Campanas Observatory, Chile.}}
   \authorrunning{Fumagalli et al.}
   \titlerunning{On the redshift of the blazar \pks}


   \author{Michele Fumagalli\inst{1},
     Amy Furniss\inst{2}, 
     John M. O'Meara\inst{3}, 
     J. Xavier Prochaska\inst{1,4},\\
     David A. Williams\inst{2},
     \and
     Emanuele P. Farina\inst{5,6}
   }
   
   \institute{Department of Astronomy and Astrophysics, University of California,  
     1156 High Street, Santa Cruz, CA 95064, USA \email{mfumagalli@ucolick.org}     
     \and
     Santa Cruz Institute for Particle Physics and Department of Physics, 
     University of California, Santa Cruz, CA 95064, USA
     \and
     Department of Chemistry and Physics, Saint Michael's College, One Winooski Park, Colchester, VT 05439, USA
     \and
     UCO/Lick Observatory, University of California, 1156 High Street, Santa Cruz, CA 95064, USA
     \and
     Universit\`a dell'Insubria, via Valleggio 11, I-22100 Como, Italy
     \and
     INFN Milano-Bicocca, Universit\`a degli Studi di Milano-Bicocca,
     Piazza della Scienza 3, I-20126, Milano, Italy
   }
   \date{Received July 16, 2012; accepted August 08, 2012}

 
  \abstract
      { 
        \pks\ is one of the BL Lacertae objects that have been detected at very high energy.
        There has been a recent report of a lower limit of $z\geq 1.246$ for the 
        redshift of this blazar, challenging the current paradigm in which 
        very high-energy $\gamma$-rays cannot freely propagate in the $z\gtrsim1$ universe.
        In this research note, we present a new MagE/Magellan spectrum of \pks\ with exquisite 
        signal-to-noise ($S/N>150$ at $6500~\AA$). Our analysis confirms the presence of the 
        previously-reported absorption line at $6280~\AA$, which we identify, however, with a known 
        telluric absorption, invalidating the claim that this blazar lies at $z>1$. Since no other 
        extragalactic spectral features are detected, we cannot establish a redshift based 
        on our spectrum.\thanks{The spectrum is available in electronic form at the CDS via 
        anonymous ftp to cdsarc.u-strasbg.fr (130.79.128.5) or via 
        http://cdsweb.u-strasbg.fr/cgi-bin/qcat?J/A+A/}}
   {}
   {}
   {}
   {}
   

   \maketitle
%

\section{Introduction}

Blazars are among the most powerful astrophysical objects so are excellent 
laboratories for studying a variety of phenomena, ranging from the origin of relativistic
particles in jets to the properties of the extragalactic background light (EBL) and 
intergalactic magnetic field (IGMF). 
Since  very high-energy (VHE; $E > 100~\rm GeV$) $\gamma-$rays are 
absorbed as they propagate through the photon field of the extragalactic background,
blazars that emit at VHE are particularly useful probes for the EBL and IGMF. Unfortunately, blazars and, 
in particular, BL Lacertae (BL Lac) objects, often exhibit 
a featureless power-law spectrum, making the task of establishing
redshifts with optical spectroscopy particularly challenging.
This nagging problem has favored the development of independent techniques
that are now being used to constrain the distance to blazars, including the 
near-IR and optical imaging of the host galaxies \citep[e.g.][]{sba05,nil08,mei10,kot11}, 
the minimum equivalent-width method to set redshift limits in featureless spectra \citep[e.g.][]{sba06}, 
the simultaneous analysis of GeV and TeV emission \citep[e.g.][]{pra12}, or  
UV and molecular spectroscopy \citep[e.g.][]{dan10,fum12}.

This research note focuses on the redshift of \pks, a BL Lac object that is detected at 
VHE \citep{zec11}. Based on weak Ca~II absorption lines, \citet{per98} report a spectroscopic 
redshift of $z=0.205$ for this source. This redshift is consistent with the lower limit $z>0.176$ 
reported by \citet{lan08} using photometric techniques, and it also agrees with the 
redshift constraints inferred from the analysis of GeV and TeV emission by  \citet{pra12} of  $z=0.2\pm0.05$ 
and by \citet{zec11} of  $z < 0.53$. However, \citet{lan12} has recently announced a much higher 
spectroscopic redshift of $z>1.246$ based on detection of Mg~II lines in two high 
signal-to-noise ($S/N$) optical spectra.
Because the mean free path of VHE $\gamma-$ray photons steeply decreases for increasing
redshift owing to electron-positron pair production against the lower
frequency EBL photons, a redshift $z>1$ for a VHE-detected blazar is quite surprising,
and it directly challenges the current paradigm for the propagation of 
VHE photons in the universe \citep[see e.g.][]{aha12}. The important implications of 
this redshift for studies of the EBL and IGMF have led us to scrutinize 
the optical spectrum of \pks further, presenting conclusive evidence 
that the previously reported redshift of $z>1.246$ is incorrect.

\begin{figure}
  \centering
  \includegraphics[scale=0.41,angle=90]{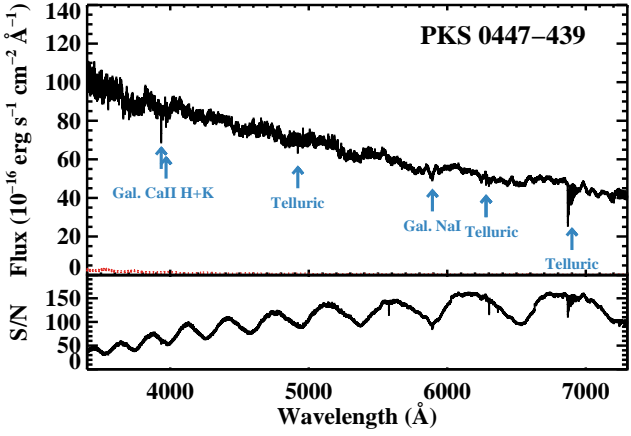}
  \caption{MagE spectrum (top) and wavelength dependent $S/N$ per pixel (bottom) for 
    the BL Lac \pks. The $1\sigma$ error is represented by the red dotted line.
    All the visible absorption lines are of either Galactic or 
    terrestrial origin, resulting in an unknown spectroscopic redshift for this source.}
  \label{spec}%
\end{figure}

\section{Observations and data reduction}

We observed the blazar \pks\ with the Magellan Echellette Spectrograph \citep[MagE;][]{mar08} mounted at 
the Clay Magellan II telescope on UT 2012 July 12. Weather conditions were fair, with mostly
clear skies and moderate seeing ($\sim 1.5''$). Two exposures of 900 seconds 
were acquired through a $0.7''$ slit. The source was observed at airmass 1.7 because  
\pks\ was transiting after the morning twilight at the time of the observations. Spectra for different
spectro-photometric standard stars were also acquired before and after the science observations.
The spectra were reduced using the \textsc{mase} pipeline \citep{boc09}, which coadds the 
1D spectra extracted from the flat-fielded and sky-subtracted 2D spectral images.
The final spectrum covers the wavelength interval $3000-10000~\AA$ with $S/N\sim50-150$ 
per pixel of 0.4~\AA. This spectrum was flux calibrated using the standard star Feige110.
Due to varying seeing conditions during the observations, the absolute flux scale 
is affected by a $\sim 20-30\%$ uncertainty. We also 
note that the wiggles superimposed on the blazar power-law are not intrinsic,
since caused by small irregularities in the sensitivity function and in the 
flat-field correction.

\begin{figure}
  \centering
  \includegraphics[scale=0.46]{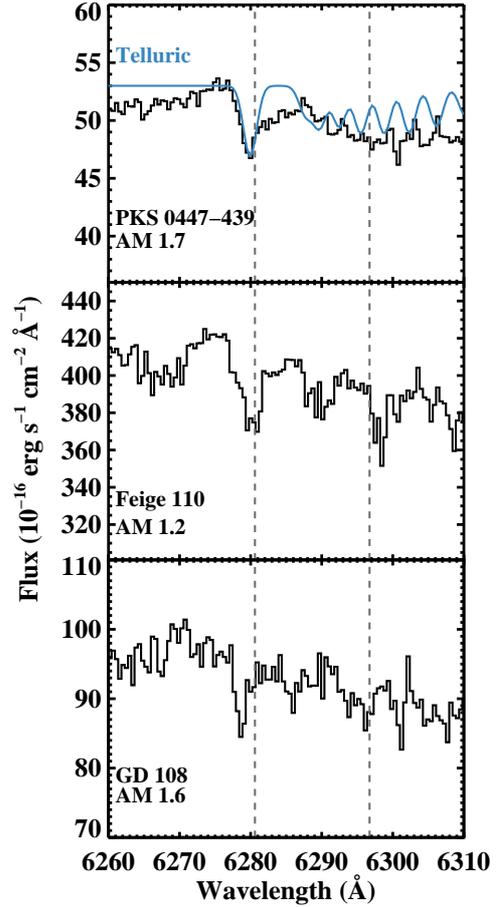}
  \caption{Zoom-in spectra for \pks\ and the two standard stars Feige~110 and GD~108
    in the wavelength interval corresponding to the claimed Mg~II absorption lines \citep{lan12}.
    In the top panel we superimpose on the spectrum of \pks\ a model of the Earth 
    atmospheric absorption (in blue), derived
    from the position and relative strength of known telluric lines (see text). 
    The two vertical dashed lines mark the frequency of the previously claimed Mg~II absorption at 
    $z=1.246$, which corresponds instead to atmospheric features.}
  \label{tell}%
\end{figure}

\section{Discussion}

\subsection{Analysis of visible spectral features}

A visual inspection of the spectrum reveals multiple 
absorption lines (see Figure \ref{spec}). The most prominent 
absorption features are telluric lines in the wavelength range $\sim 6800-6950~\AA$,
$\sim 7200-7300~\AA$, and $\sim 7600-7700~\AA$.  Also, Galactic Ca~II H+K absorption lines at
$3934.79~\AA$ and $3969.61~\AA$ are clearly detected. We further identify 
weak absorption features associated with Galactic Na~I lines at $5891.61~\AA$, 
$5894.13~\AA$, and $5897.57~\AA$. The Ca~II H+K lines found 
at $z=0.205$ by \citet{per98} are not detected in our spectrum.
Finally, two remaining absorption features are visible 
at $4921.1~\AA$ and $6280.0~\AA$.  The line at $4921.1~\AA$ is also detected 
in the spectra of multiple standard stars, proving that this is not an extragalactic feature. 

Intriguingly, the line at $6280.0~\AA$ coincides with the absorption reported by \citet{lan12}.
If associated to Mg~II at a rest-frame wavelength $2796.82~\AA$, this line would
place \pks\ at $z>1.245$, as claimed by \citet{lan12}. Neither the associated Fe~II 
absorption features (e.g. the strong $2382.76~\AA$ line) nor the Mg~I line at $2852~\AA$ are visible.
However, the same absorption line at $6280.0~\AA$ is 
also clearly detected in the spectra of multiple standard stars that have been observed at 
different airmasses throughout the observing run, confirming that this absorption is not 
associated with an intervening Mg~II systems, but originates in the Earth's atmosphere.
The independent analysis of an earlier X-shooter spectrum by \citet{pit12} 
also shows that the observed absorption is indeed atmospheric.

Figure \ref{tell} shows a comparison of the spectral region between $6260~\AA$ and $6310~\AA$ in \pks, 
Feige 110, and GD 108. Absorption lines are consistently found around $\sim 6280.0~\AA$
and lie at slightly different wavelengths, as expected from the motion of the Earth. 
Furthermore, as typically found for telluric lines, the absorption strength varies as a function
of the airmass. In Figure \ref{tell}, we show a model spectrum based on the
position and relative strength of known
O$_2$ telluric lines\footnote{http://www2.keck.hawaii.edu/inst/hires/makeewww/Atmosphere/}  
\citep[see also Fig. 1 in][]{ste94}. This spectrum has been 
shifted by $-1.8~\AA$ and arbitrarily rescaled to match the observed absorption.
The satisfactory agreement, including the evident absorption between $6290-6310~\AA$, strengthens 
our interpretation.

\subsection{Redshift lower limit}

Following the procedure described in \citet{sba05,sba06}, we set a redshift lower 
limit for \pks\ under the hypothesis that (i) the blazar host galaxy is a 
standard candle and (ii) spectral features 
cannot be detected when the ratio of the nonthermal nuclear component to the 
thermal emission from the host galaxy exceeds a maximum value set by the $S/N$ of the spectrum.
Throughout this analysis, all the values have been homogenized to modern cosmological parameters 
\citep[WMAP7;][]{kom11}.

As in \citet{sba06}, we computed the redshift-dependent nucleus-to-host ratio $\rho_{0}(z)$
for a host galaxy with absolute magnitude $M_{\rm R}=-22.9\pm0.5$  in the R-band.
For the nuclear component, we used $m_{\rm n}=13.58\pm0.15$ mag as measured from our spectrum
after accounting for slit-loss corrections, Galactic extinction, and the host galaxy contribution.  
Next, we computed the maximum nucleus-to-host ratio $\rho_{\rm 0,max}(z)$ 
for which a Ca~II ($\lambda=3934.79~\AA$) absorption line with rest-frame 
equivalent width $W_0=16~\AA$ can be detected at a given redshift. For the mean $S/N$ of our 
spectrum, the minimum detectable equivalent width is $W_{\rm min}=0.68~\AA$ \citep[see][]{sba05}.
When $\rho_0 > \rho_{\rm 0,max}$, the host galaxy spectral features are outshined  by the nonthermal 
continuum, providing a redshift lower limit of $z>0.012$ for \pks.
If we assume a fainter nuclear component of $m_{\rm n}=14.4$ from previously reported magnitudes 
we instead find $z>0.025$. Despite the high $S/N$ of this spectrum, the brightness of this blazar and the
fact that the spectrum was acquired in moderate seeing conditions meant we were prevented  
from establishing a more 
stringent limit on the redshift. 

In summary, from the analysis of this new optical spectrum of \pks, we conclude that there is no evidence 
of a redshift $z>1.24$ and that the spectroscopic distance to this blazar is only
constrained by the tentative detection of the Ca~II lines reported by \citet{per98}.

\begin{acknowledgements}
We wish to thank R. Simcoe and J. Hennawi for their assistance provided 
during the data reduction. 
\end{acknowledgements}

\end{document}